\documentclass[10pt, final, journal, letterpaper, twocolumn]{IEEEtran}

\usepackage{cite}
\usepackage{amsmath,amsthm,amssymb,amsfonts}
\usepackage{algorithm}
\usepackage{algorithmic}
\usepackage{graphicx}
\usepackage{color}
\newcommand{\bs}{\boldsymbol}
\usepackage[normalsize]{subfigure}
\usepackage{framed}

\begin{document}

\title{Data-Importance Aware Radio Resource Allocation: Wireless Communication Helps Machine Learning}
\author{Yuan Liu, Zhi Zeng, Weijun Tang, and Fangjiong Chen
\thanks{
%
The authors are with the School of Electronic and Information Engineering, South China University of Technology, Guangzhou, 510641, P. R. China (email: eeyliu@scut.edu.cn; eezhizeng@mail.scut.edu.cn; tangwj@scut.edu.cn; eefjchen@scut.edu.cn).
} 
}

\maketitle

\vspace{-1.5cm}

\begin{abstract}
The rich mobile data and edge computing enabled wireless networks motivate to deploy artificial intelligence (AI) at network edge, known as \emph{edge AI}, which integrates wireless communication and machine learning. In communication, data bits are equally important, while in machine learning some data bits are more important. Therefore we can allocate more radio resources to the more important data and allocate less radio resources to the less important data, so as to efficiently utilize the limited radio resources. To this end, how to define ``more or less important" of data is the key problem. In this article, we propose two importance criteria to differentiate data's importance based on their effects on machine learning, one for centralized edge machine learning and the other for distributed edge machine learning. Then, the corresponding radio resource allocation schemes are proposed to improve performance of machine learning. Extensive experiments are conducted for verifying the effectiveness of the proposed data-importance aware radio resource allocation schemes.
\end{abstract}


\section{Introduction}

In recent years, we have witnessed the explosive growth in mobile data,  most of which are generated by wireless devices (WDs), like smartphones and internet-of-things (IoT) sensors. The massive data are usually uploaded to cloud center for training artificial intelligent (AI) models. However,  traditional cloud AI suffers from network congestion and does not support real-time applications. Mobile-edge computing (MEC), placing cloud-like functions at the network edge [like base station (BS) or access point (AP)], is an emerging technology that can overcome the shortcomings of cloud technology \cite{liu_zhang,MengyuWCL,latence_mec,Lzz}.

Moreover, MEC further enables \emph{edge AI} or \emph{edge machine learning} that employs machine learning at edge servers, which integrates wireless communication and machine learning. The integration can be divided into two directions: the first is machine learning for wireless communication (ML-WC) and the second is wireless communication for machine learning (WC-ML). Most of the existing works can be categorized into ML-WC (please see survey papers in \cite{CXMCM,Osvaldo,Park,Jithin,StrinatiVTM,QinWCL,JWang,ZhangCST,ChenCST} and references therein), i.e.,  using machine learning as a tool to solve  very complex problems or resolve mathematically intractable expressions pertinent to
wireless communications. For WC-ML, edge servers use data transmitted from WDs to train machine learning models; in return, WDs download the trained models to quickly respond to real-time events. As the speed of data processing and computing can be rapidly increased at edge servers, wireless communication becomes a bottleneck for fast learning in WC-ML, since training data are usually large in scale but radio resources are scarce. In this case, machine learning operations and performance hinge on radio resources and channel dynamics. This calls for efficient radio resource allocation for fast learning and opens up a new research paradigm that is largely uncharted so far. Accordingly, this motivates us to consider how wireless communication can help machine learning at edge.

%

In conventional communication systems, the goal is for either reliable transmission or maximizing data rates, in which data bits are equally important. This can simplify the system design. However, it can not explore the feature of machine learning because some data bits are more important than others in machine learning, e.g., the data near the decision boundary of a classifier are more important than those far away \cite{Settles}. Motivated by this,  in \cite{DongSPAWC19}, the distance between the data to the decision boundary of support vector machine (SVM) classifier is characterized by signal-to-noise ratio (SNR) of the data.

\begin{figure*}
\centering
\includegraphics[width=11cm]{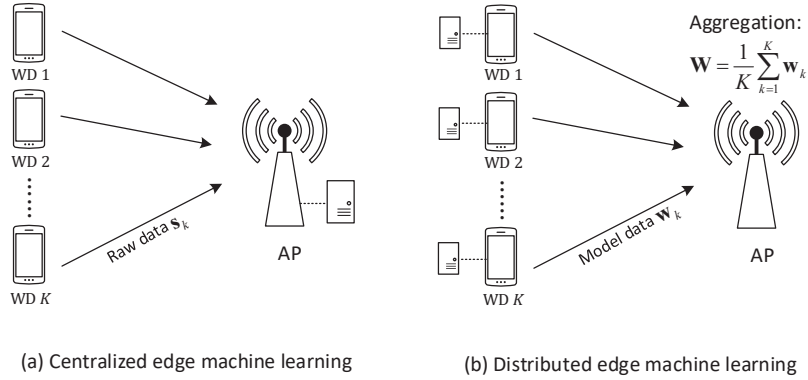}
\caption{System models of edge machine learning.}
\label{fig:model}
\end{figure*}

In this article, we consider an edge machine learning system as shown in Fig. \ref{fig:model}, consisting one access point (AP) and multiple WDs. A certain machine learning model is trained at the AP by using the data transmitted from the WDs. We propose two simple and easily implemented importance criteria to differentiate the WDs' data-importance, with one for centralized edge machine learning and the other for distributed edge machine learning. As a result, more radio resources are allocated to more important data to combat noise, targeting accelerating convergence and improving accuracy of machine learning models. We evaluate the proposed schemes using real datasets and show that performance gain can be achieved compared with the traditional schemes.
%
%
Note that our goal is to propose simple and easily implementable importance criteria in which we show that even simple communication design (retransmission used in this article for example) can improve the performance of machine learning. Definitely other existing sophisticated communication designs under the proposed importance criteria can further improve learning performance but this is beyond the scope of this article.

The remainder of this article is organized as follows. Section \ref{sec:central} and Section \ref{sec:distr} detail the proposed importance criteria for the centralized and distributed edge machine learning systems, respectively. Section \ref{sec:sim} presents experimental results for the proposed schemes. Section \ref{sec:con} finally concludes this article.

\section{Centralized Edge Machine Learning}\label{sec:central}

In this section, we consider a centralized edge machine learning system as shown in Fig. \ref{fig:model}(a), where the machine learning model is trained at the AP by using samples received from $K$ WDs. Consider supervised learning in this paper, $\bs s_i:= (\bs d_i, \ell_i)$ is denoted as the sample (from any WD $k$), where $\bs  d_i$ is the data and $\ell_i$ is the corresponding label. Note that a data $ \bs d_i$ usually has much larger size (with million coefficients for example) than a label $\ell_i$  (with a $0\sim 9$ integer for example). Thus it is assumed that the WDs' labels can be correctly received at the AP via a noiseless \emph{label channel},\footnote{``Noiseless" means that the channel is without noise or the noise can be neglected. } while the WDs' data  are transmitted over the noisy and fading \emph{data channel}. The data channel is assumed to be block fading so that the channel gains remain unchanged during each resource block but vary from one resource block to another.  Let $N$ denote the maximum number of resource blocks, each of which is used for transmitting one training sample. The additive white Gaussian noise (AWGN) at the data channel is assumed to be independent circular symmetric complex Gaussian random process.  Moreover, we assume that global channel state information (CSI) are available at the AP. 

In machine learning, the training data are required to be correct as wrong data lead to incorrect learning. Thus when receiving a sample $\bs s_i$, the AP needs to check whether the data $\bs d_i$ matches its corresponding label $\ell_i$ or not under the current model denoted $\bs W_{i-1}$. We say that the received data $\bs d_i$ is correctly classified if $\bs d_i$ matches $\ell_i$, denoted as $\bs W_{i-1}(\bs d_i)=\ell_i$,  and the received data $\bs d_i$ is wrongly classified otherwise, denoted as $\bs W_{i-1}(\bs d_i)\neq\ell_i$. For each wrongly classified data, the reasons are two-folds: one is that the data channel is too noisy to receive the data, and the other is that the current model $\bs W_{i-1}$ itself is wrong. Due to these reasons, we define the importance criterion of centralized edge machine learning as follows:
%
\begin{framed}
\noindent
\textbf{Importance Criterion 1}. Wrongly classified data are more important and correctly classified data are less important.
\end{framed}
The intuition behind the aforementioned criterion is that, for a given training data, if the current model's judgement on the data does not match its label, it means that the current model needs to be adjusted and the data is important to such adjustment. On the contrary, if the current model's judgement on the data matches its label, it means that the current model does not need to be adjusted to some extent and thereby the data is less important in this case.

To this end, we set higher received SNR threshold $\overline{\gamma}$ for the more important data and lower received SNR threshold $\underline{\gamma}$ for the less important data,  i.e., 
\begin{eqnarray}\label{eqn:thr}
\textsf{SNR}(\bs d_k)=\begin{cases}\overline{\gamma},~\text{if $\bs W_{i-1}(\bs d_i)\neq\ell_i$}\\
\underline{\gamma},~\text{if $\bs W_{i-1}(\bs d_i)=\ell_i$,}\end{cases}
\end{eqnarray}
where $\overline{\gamma}>\underline{\gamma}$ and they are system parameters and constants. It is noticed that $\overline{\gamma}=\underline{\gamma}$ in conventional communication systems where data bits are equally important.

According to the importance criterion defined above, we adopt the classic automatic repeat-request (ARQ) policy to allocate resource blocks. That is, after determining the data importance, each data $\bs d_i$ is allocated a new resource block for retransmission. By maximal ratio combining (MRC) at the AP, the retransmission is repeated until the received SNR exceeds the threshold defined in \eqref{eqn:thr}.  Here the goal of retransmission is to suppress fading and noise so as to increase data reliability and thus  learning performance. As $\overline{\gamma}>\underline{\gamma}$ is set in \eqref{eqn:thr}, the more important data are allocated more resource blocks than the less important data to improve learning performance.
After retransmission, each data $\bs d_i$ together with the previous received data, i.e., $(\boldsymbol d_1,\cdots,\boldsymbol d_i)$, are used to update the current model $\boldsymbol W_{i-1}$ into a new model $\boldsymbol W_{i}$. 

Finally, we depict the training procedure for the centralized edge machine learning in Algorithm 1, which consists of three steps: data judgement, data retransmission, and model update. The training procedure ends if the model converges or the totally $N$ resource blocks are exhausted in the retransmission step.

\begin{algorithm}[!t]
\caption{Training for centralized edge machine learning}
\begin{algorithmic}[1]
\STATE \textbf{initialize} $\boldsymbol W_0$ and $i=0$.
\REPEAT 
\STATE $i\leftarrow i+1$;
\STATE \emph{(Data Judgement):} Data $\boldsymbol d_i$ is more important if $\boldsymbol W_{i-1}(\boldsymbol d_i)\neq\ell_i$, and less important otherwise;
\STATE \emph{(Data Retransmission):} Retransmit $\boldsymbol d_i$ using a new resource block and combine $\boldsymbol d_i$ by MRC at AP until the received SNR exceeds the threshold \eqref{eqn:thr};
\STATE \emph{(Model Update):} Update model $\boldsymbol W_{i-1}$ by using data $(\boldsymbol d_1,\cdots,\boldsymbol d_i)$ into a new model $\boldsymbol W_{i}$;
 \UNTIL{Model converges or $N$ resource blocks exhaust. } 
\end{algorithmic}
\end{algorithm}

\section{Distributed Edge Machine Learning}\label{sec:distr}

We now study the distributed edge machine learning case as shown in \ref{fig:model}(b). Different from the centralized edge machine learning where the learning task is embedded at the AP while the WDs transmit raw data, the distributed edge machine learning distributes the learning process over the WDs, and the WDs transmit their individual local models trained by their local datasets, then the AP aggregates the local models for the global model.

Denote $D_k=|\bs d_k|$ is the size of dataset of WD $k$. The distributed edge machine learning framework is that, each WD transmits the trained model $\bs w_k$ to the AP instead of raw data $\bs d_k$ and, at the AP side, the global model is obtained by aggregation or averaging:
\begin{equation}\label{eqn:agg}
\bs W=\frac{1}{K}\sum_{k=1}^K \bs w_k.
\end{equation}
 For above traditional distributed learning framework expressed via \eqref{eqn:agg}, the local models $\{\bs w_k\}$ have equal importance.

Note that the local models $\{\bs w_k\}$ also experience independent fading and noise when they are transmitted to the AP. On the other hand, the local model trained by using  larger size of dataset in general has higher accuracy in machine learning. Based on this, we distinguish the data importance of the local models and define the importance criterion of distributed edge machine learning as follows:
\begin{framed}
\noindent
\textbf{Importance Criterion 2}. The data of local model trained by larger dataset are more important and that by smaller dataset are less important.
\end{framed}

Therefore, following by the idea of more resource blocks being allocated to more important data, we adopt the widely used proportional allocation of resource blocks to the WDs. Specifically, each WD $k$ is allocated $N_k$ resource blocks for retransmission of model data $\bs w_k$, proportionally to its dataset size $D_k$, i.e.,
\begin{equation}
N_k = \left\lfloor\frac{D_k}{\sum_{k=1}^K D_k} N\right\rfloor,
\end{equation}
where $N$ is the total number of resource blocks and $\lfloor \cdot \rfloor$ is the floor operator. Denote $\bs w_k(n)$ as the $n$-th retransmission of local model $\bs w_k$, the \emph{data-importance aware} aggregation is
\begin{equation}\label{eqn:agg1}
\bs W'=\frac{1}{N}\sum_{k=1}^K\sum_{n=1}^{N_k} \bs w_k(n).
\end{equation}
In another word, each local model $\bs w_k$ is retransmitted $N_k$ times. After retransmission of each local model, all the data copies of $\{\bs w_k\}$ are aggregated as in \eqref{eqn:agg1} for the global model. Compared with the traditional aggregation with equal importance in \eqref{eqn:agg}, the effects of retransmission here are two-fold: one is to using more resources to suppress fading and noise for more importance local models for increasing data reliability like the centralized case, and the other is to increase the proportion of the more important local models in aggregation. Both  the effects finally improve  performance of machine learning algorithms.

\section{Experimental Results}\label{sec:sim}

In this section, we evaluate the proposed data-importance criteria via simulations. The wireless fading are assumed to be independent and identically distributed (i.i.d.) Rayleigh fading. A resource block  corresponds to a time-slot.  For a fair comparison with \cite{DongSPAWC19}, the AP adopts SVM as the machine learning algorithm for the centralized edge machine learning, while for the distributed edge machine learning, all the WDs also use the SVM model. And the SVM model uses linear kernel with the default parameters.  We use the well-known mixed national institute of standards and technology  (MNIST) dataset of handwritten digits to train the SVM classifier, which consists of two parts: a training set containing 60,000 samples and a test set containing 10,000 samples, and each set comprises data and labels. Each data in the MNIST data set is a gray image of 28$\times$28 pixels, which means the dimension of a data is 784, corresponding to 784 columns in the data matrix, and each row is a gray image. The content of these data is handwritten numbers 0$\sim$9, and these ten categories corresponds to the 10 columns of the label matrix, while each row represents the corresponding image located in the same row of the data matrix. In every row, only one column that the category belongs to is marked as 1, and the others are marked as 0. After training the classifier by using the training set, the test set is used to evaluate the classifier's accuracy, which is similar to data judgement step in Algorithm 1. Specifically, the trained classifier judges each data in test set, if the result matches its label, the judgement is correct and wrong otherwise. The final classifier's accuracy is averaged over all data in test set.

\subsection{Centralized Edge Machine Learning}

We consider two benchmarks in centralized edge machine learning case, i.e., in this subsection Benchmark 1 denotes the scheme of \cite{DongSPAWC19} and Benchmark 2 denotes the traditional scheme with equal importance.
To see the best performance of the three schemes, we exhaustively search the parameters, i.e., the values of received SNR thresholds ($\gamma$ for traditional scheme,  $\overline\gamma$ and $\underline\gamma$ for the proposed scheme), and the value of data-alignment probability in \cite{DongSPAWC19}, then we choose the parameters achieving the best performance for all schemes for fair comparison.
It is worth noting that in practical communication systems, the SNR thresholds are preset values depending on specific applications/scenarios and do not need to search when run the algorithms. In another word, the SNR thresholds are preset system parameters rather than optimization/searching variables.

\begin{figure}
\centering
\includegraphics[width=9.5cm]{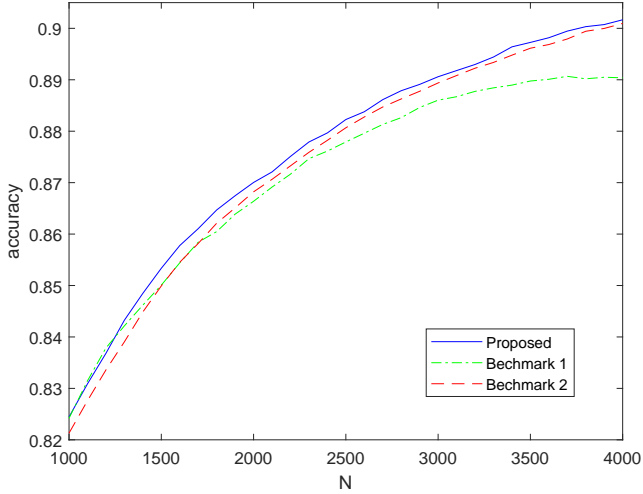}
\caption{Accuracy of centralized machine learning versus number of resource blocks $N$.}
\label{fig:threeN}
\end{figure}

In Fig. \ref{fig:threeN}, we study the accuracy of the trained model versus the number of resource blocks $N$, where the transmitted SNR is fixed as 4 dB for all schemes. We can observe that the performance of all schemes improve when the number of resource blocks increases. This is expected since more available resource blocks are more beneficial to suppress noise and thus improve quality of the received training data. It is also observed that the proposed scheme is superior to the two benchmark schemes, and Benchmark 1 is better than Benchmark 2 at small $N$ and then worse at large $N$.

\begin{figure}
\centering
\includegraphics[width=9.5cm]{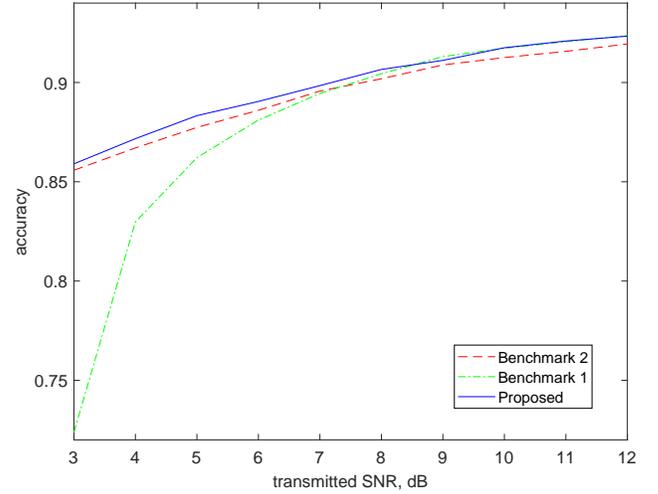}
\caption{Accuracy of centralized machine learning versus transmitted SNR of WDs.}
\label{fig:threeSNR}
\end{figure}

The accuracy of the trained model versus the transmitted SNR is investigated in Fig. \ref{fig:threeSNR}, where the number of resource blocks is fixed as $N=4000$. We observe that increasing transmitted SNR of WDs can also improve quality of training data and thus the learning performance. It also demonstrates that the proposed scheme have better learning performance than the two benchmark schemes. Moreover, the proposed scheme and Benchmark 1 achieve about the same performance at high SNR regime, e.g., greater than about 7 dB in our simulation. 

\begin{figure}
\centering
\includegraphics[width=9.5cm]{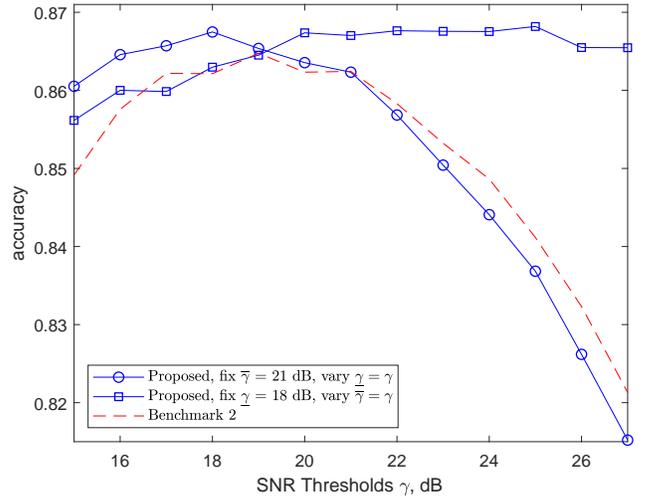}
\caption{Accuracy of centralized machine learning versus SNR threshold $\gamma$.}
\label{fig:threshold}
\end{figure}

We study the accuracy of the trained model versus the SNR thresholds in Fig. \ref{fig:threshold}, where we set the number of resource blocks as $N=2000$. For the proposed scheme, we fix one of $\overline\gamma$ and $\underline\gamma$ and vary the other one as $\gamma$ of the traditional scheme with equal importance (i.e., Benchmark 2). We can observe that, as $\gamma$ increases, all three curves first increase and then decrease. This means that there always exists one optimal SNR threshold achieving best performance for each scheme. Moreover, it shows that the proposed scheme can be always better than the traditional equal-importance scheme by selecting proper SNR thresholds.

\subsection{Distributed Edge Machine Learning}

We also consider two benchmark schemes for distributed edge machine learning. The first benchmark scheme is equal-importance scheme, i.e., each WD is equally allocated $N/K$ resource blocks (note again that $N$ and $K$ are the numbers of resource blocks and WDs, respectively). The second one is that only the WD with the largest dataset is chosen to upload its local model. Here total $N=200$ resource blocks are considered and the SNR is fixed as 20 dB.

\begin{figure}
\centering
\includegraphics[width=9.5cm]{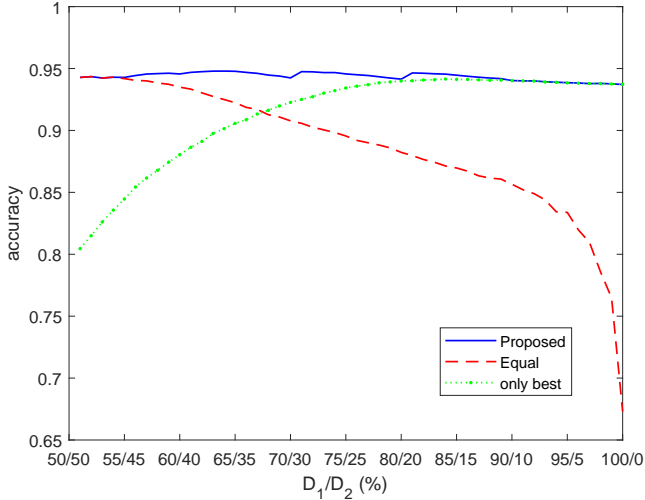}
\caption{Accuracy of distributed machine learning versus ratio of dataset sizes $D_1/D_2$.}
\label{fig:disD}
\end{figure}

We first consider two WDs with different dataset sizes (i.e., $D_1$ and $D_2$) impacting on the accuracy of the trained model, as shown in Fig. \ref{fig:disD}, where the whole dataset is split into two disjoint sub-datasets with each for one WD. Firstly we can observe that the proposed scheme is superior to the two benchmark schemes over all ratios of $D_1/D_2$. In addition, we observe that the accuracy of the proposed scheme is slightly decreasing when the dataset sizes becomes imbalanced. This indicates that the proposed scheme has best performance of resistance against the imbalance of dataset sizes. The equal-importance scheme becomes worse when the dataset sizes are more imbalanced. This interestingly reveals that, in traditional scheme, equal treatment of all local models will lower accuracy of the global model. On the contrary, the accuracy of the scheme selecting the largest dataset is increasing with data imbalance, which achieves about the same performance with the proposed scheme when one WD has all data.

\begin{figure}
\centering
\includegraphics[width=9.5cm]{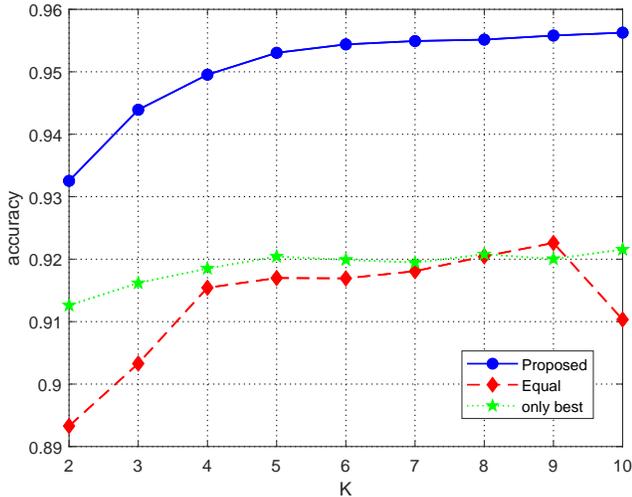}
\caption{Accuracy of distributed machine learning versus  number of WDs $K$.}
\label{fig:disK}
\end{figure}

Then we study the impact of number of WDs $K$ on accuracy of the global model in Fig. \ref{fig:disK}, where the whole dataset is randomly split into $K$ disjoint sub-datasets, with each sub-dataset for one WD. The superiority of the proposed scheme is again validated over the two benchmark schemes. For the proposed scheme, the accuracy of the global model is improved when the participated WDs increase, even though the total used data are fixed as the whole dataset is given. This is because that the dataset sizes of WDs become more balanced when the number of WDs becomes large as the dataset sizes are randomly generated from the given whole dataset.

\section{Concluding Remarks}\label{sec:con}

In this article, we proposed two new data-importance criteria for mobile data transmission in centralized and distributed edge machine learning, respectively. The idea was to differentiate mobile data‘s importance according to their effects in machine learning, and radio resources are accordingly allocated to improve data quality and thus machine learning performance. We showed that, under the proposed data-importance criteria, simple radio resource allocation schemes can efficiently improve performance of machine learning.

\bibliographystyle{IEEEtran}
\bibliography{IEEEabrv,CMEC}

\end{document}